# Understanding the Effects of Dielectric Property, Separation Distance, and Band Alignment on Interlayer Excitons in 2D Hybrid MoS$_2$/WSe$_2$ Heterostructures

Jaehoon Ji and Jong Hyun Choi*
School of Mechanical Engineering, Purdue University
West Lafayette, Indiana 47907, United States
*Corresponding author: jchoi@purdue.edu

**Abstract**
Two dimensional (2D) van der Waals heterostructures from transition metal dichalcogenide (TMDC) semiconductors show a new class of spatially separate excitons with extraordinary properties. The interlayer excitons ($X_I$) have been studied extensively, yet the mechanisms that modulate $X_I$ are still not well understood. Here, we introduce several organic-layer-embedded hybrid heterostructures, MoS$_2$/organic/WSe$_2$, to study the binding energy of $X_I$. We discover that the dielectric screening of the quasi-particle is reduced with organic molecules due to decreased dielectric constant and greater separation distance between the TMDC layers. As a result, a distinct blueshift is observed in interlayer emission. We also find that the band alignment at the heterointerface is critical. When the organic layer provides a staggered energy state, interlayer charge transfer can transition from tunneling to band-assisted transfer, further increasing $X_I$ emission energies due to a stronger dipolar interaction. The formation of $X_I$ may also be significantly suppressed with electron or hole trapping molecules. These findings should be useful in realizing $X_I$-based optoelectronics.

Two dimensional (2D) heterostructures built by stacking transition metal dichalcogenides (TMDCs) such as $MoS_2$ and $WSe_2$ have garnered much attention due to their extraordinary optoelectronic properties. The strong light-matter interaction in the heterostructures gives rise to interlayer excitons ($X_I$) which are a bound state of spatially separate electrons and holes in opposite 2D layers.[1] The $X_I$ in heterobilayers such as $MoS_2/WSe_2$ and $MoSe_2/WSe_2$ have large binding energies (~110 meV).[2-4] The large energy leads to stable photoluminescence (PL) emission with a long lifetime (~100 ns)[5] and scalable diffusion length (~a few μm)[6]. These characteristics make TMDC heterostructures suitable for light-emitting diodes,[7] photodetectors,[8] and photovoltaics.[9] In addition, the separation of charges brings about permanent electric dipoles in the out-of-plane direction of the structure.[10] This allows for electrical control of the interlayer excitons, which may not be possible with individual TMDC layers.[11] The exceptional properties and controllability could be harnessed to realize emerging applications such as excitonic devices and valleytronics.[12]

The interlayer excitons in TMDC heterostructures are formed by tunneling of the photogenerated charges through the van der Waals gap between the stacked TMDC heterolayers.[13] Various mechanisms including chemical doping,[2] external electrical field,[10] and defects[14-15] were investigated to understand the formation and recombination of $X_I$. For example, the lifetime of $X_I$ in 2D TMDC heterostructures may be increased by an order of magnitude by photo-irradiation and gate voltage.[11] The $X_I$ recombination energy can be also highly tuned (*e.g.*, redshift up to ~100 meV) by applying electrical potential due to the Stark effect.[16] The defects in the heterolayers may allow localized interlayer exciton-polariton interactions, resulting in the enhanced interlayer emission.[14] Despite the extensive studies, understanding the mechanisms that govern $X_I$ and the related emission still remains limited. For example, the dielectric property can change the binding energy of $X_I$ by disrupting the Coulombic interaction between the charges (*i.e.*, electron-hole or e-h). Previous studies mostly focused on the dielectric property of the environment (*e.g.*, various substrates),[17-18] while few studies investigated that of the heterolayers.[19-20] Modulating the dielectric properties of the heterostructures, for example, by inserting a layer between TMDC layers, may provide critical insights and novel strategies for tailoring the $X_I$ emission signatures. Further, the insertion of an intermediate layer will necessarily change the separation distance between the photoinduced e-h charges, affecting the $X_I$ characteristics. Such an approach could also help study additional mechanisms that can tailor the interlayer charge transfer process and thus regulate the $X_I$ energy states. The new level of understanding would form the foundation for $X_I$-based applications.

This work elucidates the dielectric property, the separation gap, and the charge transfer mechanisms between the stacked heterolayers by embedding a uniform layer of organic molecules between them. The 2D $MoS_2$/organic/$WSe_2$ hybrid heterostructures show distinct characteristics compared with $MoS_2/WSe_2$, and the overall properties depend strongly on the nature of the organic molecules. We discover that the dielectric constant of the hybrid heterostructures is reduced by the inserted molecules such as 1,3-bis(3,5-dipyrid-3-ylphenyl)benzene or B3PyPB which has a large energy gap between highest occupied molecular orbital (HOMO) and lowest unoccupied molecular orbital (LUMO). The organic layer also increases the distance between e-h charges in $X_I$. As a result, the organic-layer-embedded structure exhibits a strong blueshift in the $X_I$ emission. Besides the dielectric screening, we find that the band alignment with the organic layer can also change the interlayer emission drastically. Eosin Y (EY) forms an energy state between $MoS_2$ and $WSe_2$, thereby promoting a transition from tunneling to band-assisted transport. Consequently, even higher energy emission (*i.e.*, blueshift in $X_I$ PL) is observed. Further, the interlayer emission may also be suppressed completely when electron or hole trapping layers are used including tetracyanoquinodimethane (TCNQ) and cobalt phthalocyanine (CoPc), respectively. These findings provide critical insights on the interlayer excitons and should be useful in designing optoelectronic devices from 2D TMDC materials.

To investigate the effects of distinct mechanisms on $X_I$, we prepared the organic-molecule-embedded hybrid heterostructures with a sequential stacking method (see Materials and Methods for details).[2, 21] First, monolayer $MoS_2$ flakes were placed on a silicon ($Si/SiO_2$) substrate by mechanical exfoliation.[22] The $MoS_2$ samples were subsequently immersed in the solutions containing dye molecules to form organic layers on top of $MoS_2$ flakes. Atomic force microscopy (AFM) imaging reveals that the organic molecules, B3PyPB, EY, TCNQ, and CoPc, form uniform layers with an almost identical thickness of ~0.9 nm (Figures S1 and S3; also see SI for details). Then, monolayer $WSe_2$ was transferred onto the $MoS_2$/organic sample by polydimethylsiloxane (PDMS) stamping.[2] $MoS_2/WSe_2$ heterobilayers were also prepared as a control with the same method except for the solution-incubation step.

The irradiation from a HeNe laser at 633 nm generates interlayer excitons as well as excitons in individual $MoS_2$ and $WSe_2$ layers as illustrated in Figure 1a. The staggered energy gap between $MoS_2$ and $WSe_2$ facilitates the dissociation of the excitons in each flake, and the photoexcited electrons in $WSe_2$ and holes in $MoS_2$ can transfer to the opposite layers via tunneling (Figure 1b).[23] The recombination of the spatially separate electron-hole pair leads to the interlayer emission. The organic layer embedded between $MoS_2/WSe_2$ can affect the formation of the interlayer excitons and the relevant emission. For example, B3PyPB has a large gap between its HOMO and LUMO levels (~4 eV).[24] As such, the tunneling will be the sole mechanism for photoinduced charge transfer in $MoS_2$/B3PyPB/$WSe_2$ (as illustrated in Figure 1c), in the same way with the control $MoS_2/WSe_2$. However, the insertion of an intermediate layer will increase the distance between the TMDC layers and lower the effective dielectric constant of the structure.[25] As a result, the energy states of $X_I$ and related emission will be significantly impacted (*vide infra*).

The organic molecules may also alter the band alignment, thereby causing different charge transfer pathways. EY has a LUMO situated between the CBM of $MoS_2$ and $WSe_2$.[2] As a result, the EY layer will favor the band-assisted transport of photoexcited electrons, while the holes will transfer via tunneling (Figure 1d). Other mechanisms may also be explored such as trapping of the electrons from both TMDC layers by implementing molecules with a LUMO level lower than the $MoS_2$ CBM. TCNQ is such an exemplary molecule with its LUMO at approximately -4.7 eV. Similarly, one can design hole trapping by constructing a layer of molecules such as CoPc whose HOMO (~-4.9 eV) is higher than the VBM of both $MoS_2$ and $WSe_2$. It is worth noting that all four molecular layers have a uniform thickness, and thus the effect of the extended distance between $MoS_2$ and $WSe_2$ monolayers will be similar.

Figure 2 shows the effects of the enhanced dielectric constant and the expanded separation distance in the B3PyPB-inserted $MoS_2/WSe_2$ heterostructure. The control $MoS_2/WSe_2$ sample exhibits two distinct peaks in the PL spectrum (Figure 2a). The spectral deconvolution with three Gaussian functions (see Figure S2) reveals the interlayer emission at ~801 nm as well as $MoS_2$ and $WSe_2$ PL at approximately 668 and 750 nm, respectively.[2, 26] It is notable that the interlayer emission is stronger than the intralayer PL, which may be attributed to the strong coupling of the spatially separate e-h charges in the TMDC layers, as illustrated in Figure 2b.[23] The sharp features at short wavelengths (around 650 nm) correspond to the Raman scattering of the heterostructure.[2] In Figure 2c, $MoS_2$/B3PyPB/$WSe_2$ displays several distinct features as well as similar characteristics. Its intralayer PL peaks are measured at approximately 665 nm (MoS$_2$) and 746 nm (WSe$_2$) which are slightly blueshifted (by about 3 nm) from the control sample. Figure S4 also presents the PL spectra of individual TMDC flakes with and without organic layers, for a comparison. The excitonic emission energy ($E_{PL}$) is determined by the bandgap of the material ($E_G$) and the binding energy of the exciton ($E_B$): $E_{PL} = E_G - E_B$.[27] The nearly invariable energies of intralayer PL emissions from the constituent MoS$_2$ and WSe$_2$ indicate that the extent of bandgap renormalization ($\Delta E_G$) due to the insertion of the organic layer is comparable to the change in the binding energies of intralayer excitons ($\Delta E_B$) which may not be significant.[28-30] In contrast, the interlayer emission appears as a shoulder at ~776 nm which is significantly blueshifted from the control by ~25 nm or ~50 meV. The peak intensity of interlayer emission significantly drops, while that of WSe$_2$ PL increases by order of magnitude. With the unchanged bandgaps of TMDC layers, the blueshift in $X_I$ emission may be attributed to the reduction of the binding energy of interlayer excitons.

To understand the observed phenomena, we used a simple hydrogen-like particle model to define the energy state of $X_I$.[31] Here, we take into account both the dielectric constant of the heterostructure and the separation distance between electron and hole in $X_I$ (see the SI for details). The Hamiltonian (H$_{ex}$) of the quasi-particle may be expressed as:[32-33]

$$H_{ex} = -\frac{\hbar}{2m_{ex}}\frac{1}{\rho}\frac{d}{d\rho}\left[\rho\frac{d}{d\rho}\right] - \frac{e^2}{\varepsilon_{ex}\sqrt{\rho^2+l^2}} \quad (1)$$

where $\rho$ is the relative coordinate, $e$ is an electrical charge, and $m_{ex}$ denotes the effective mass of $X_I$ as $m_{ex}^{-1} = m_e^{-1} + m_h^{-1}$, where $m_e$ and $m_h$ are the mass of electron and hole, respectively, and $\hbar$ is Planck constant. $l$ is the separation length between the charges, and $\varepsilon_{ex}$ indicates the effective dielectric function of the heterolayers. The eigenvalue of H$_{ex}$, that is, the binding energy $E_B$ of $X_I$, is closely related to the shift of interlayer emission and varies as a function of $l$ and $\varepsilon_{ex}$. To estimate $\varepsilon_{ex}$, we use a simple dielectric model (details shown in the SI Section 5) where each layer is considered as the piecewise dielectric as shown in Figure 2e.[34-36] Here, the organic layer is the additional, intermediate layer between TMDCs in the hybrid structure. The calculated dielectric function of MoS$_2$/organic/WSe$_2$ is $\varepsilon_{ex}$= ~5.4, which is smaller than that of the control (~5.7). The dielectric function can yield the Bohr radius of $X_I$ ($a = \varepsilon_{ex}\hbar/m_{ex}e^2$):[36-37] approximately 1.17 nm for the hybrid and 1.23 nm for the control. The $X_I$ binding energy can then be expressed as a function of the separation distance and Bohr radius or the distance ratio ($x = l/a$), as shown in Figure 2f. The AFM height profiles of the samples in Figures S1 and S3 reveal that the organic-layer-inserted hybrid structures have consistently greater heights than control MoS$_2$/WSe$_2$ by ~0.9 nm. With knowledge of the distance and Bohr radius in each sample, we estimate $X_I$ binding energies of the hybrid structure (~81 meV) and the control (~116 meV). The results strongly suggest that the blueshift of interlayer emission (~50 meV) in MoS$_2$/B3PyPB/WSe$_2$ may arise from the significant reduction of $X_I$ binding energy (~35 meV) and the renormalized bandgap[38] due to the expanded distance and the reduced screening.

Next, we explored EY whose LUMO level is between CBM of WSe$_2$ and MoS$_2$. Figure 3a shows the PL spectrum of the MoS$_2$/EY/WSe$_2$ heterostructure. Three distinct PL signatures are observed. The intensity and peak positions of MoS$_2$ (~664 nm) and WSe$_2$ (~746 nm) are nearly identical with those from MoS$_2$/B3PyPB/WSe$_2$. The interlayer emission, however, is measured at around 764 nm, which is even further blueshifted from MoS$_2$/B3PyPB/WSe$_2$ by ~12 nm (or ~25 meV). The emission intensity is also stronger by a factor of two. This observation suggests that besides the dielectric screening[39], other mechanisms may impact in the energy state of $X_I$ in the EY-inserted heterostructure. Unlike B3PyPB, the EY layer may favor the band-assisted transfer of photoexcited electrons from WSe$_2$ to MoS$_2$ due to the staggered band alignment (Figure 3b). As a result, a greater amount of electrons will be transferred in the EY-embedded hybrid structure. Further, the energy barrier between MoS$_2$ VBM and EY HOMO is much smaller than that between MoS$_2$ VBM and B3PyPB HOMO, thus promoting more hole transfer. With the enhanced charge transfer, MoS$_2$/EY/WSe$_2$ may have a greater population of electrons in MoS$_2$ and holes in WSe$_2$. The EY-inserted heterostructure will thus have a stronger dipole-dipole interaction than the B3PyPb-embedded sample. We attribute this strong dipolar interaction to the origin of the additional blueshift observed in the MoS$_2$/EY/WSe$_2$ structure.

The interaction energy ($U$) originating from the dipole-dipole interaction may be expressed as:[40]

$$U = -\frac{e^2}{\varepsilon_{ex}\sqrt{\rho^2 + l^2}} + \frac{2e^2}{\varepsilon_{ex}\rho} . \qquad (3)$$

where the first term on the right represents the attractive interaction inside the $X_I$ and the second term stands for the repulsive dipolar interaction between $X_I$. From the mean-field approximation where the excitons follow the Boltzmann distribution,[41] the average interaction energy of the dipoles ($E_{dip}$) may be evaluated as

$$E_{dip} = n \int U(\rho) g(\rho) d\rho \qquad (4)$$

where $n$ is the dipole density. $g(\rho)$ represents the density distribution of the pair correlation: $g(\rho) = exp[-U(\rho)/k_BT]$, where $T$ is the temperature and $k_B$ is the Boltzmann constant. The difference in the interaction energy ($\Delta E_{dip}$) responsible for the PL blueshift may be written as:[40]

$$\Delta E_{dip} = \frac{4\pi e^2 l}{\varepsilon_{ex}} D\left(\frac{\varepsilon_{ex} l k_B T}{e^2} = z\right) \Delta n \qquad (5)$$

where $D$ is a bias function with respect to a dimensionless parameter $z$ which is a ratio between thermal and electrostatic energies of the charges ($z \sim 0.13$, in our system). For $z < 1$, $D$ is reduced to a form $D = \frac{\Gamma(4/3)}{2} z^{1/3}$, where $\Gamma$ is a gamma function.[40] The model suggests that $\Delta E_{dip}$ has a linear relationship with $\Delta n$, from which we obtain $\Delta n \sim 2.7 \times 10^{12}$ cm$^{-2}$. This is the density difference of photogenerated charges between EY- and B3PyPB-emebedded heterostructures (filled red circle in Figure 3d). We recently reported an increased population of the charged particles in hybrid structures, where TMDCs form a staggered band alignment with the organic layers.[11, 42] With the band-assisted transport at the WSe$_2$/EY interface, EY-functionalized hybrid structures demonstrate significantly higher electrical currents and photoinduced currents than the control samples without an EY layer. In addition, other studies also reported the energy shift of interlayer emission due to the $X_I$ density difference. For example, electrical gating can change the $X_I$ population and demonstrate a blueshift of ~20 meV in the interlayer emission,[10] which is consistent with our observation in Figure 3. Laser irradiation also shows a similar correlation between $X_I$ population and emission energy in 2D heterostructures.[43] The results strongly support that the interlayer emission can be tailored by not only dielectric screening, but also dipolar interaction due to the change in the $X_I$ population with the insertion of an organic layer in the TDMC heterolayers.

We also studied the trapping of photoexcited electrons and holes from the heterostructures, including TCNQ and CoPc layers. Figures 4a and 4b show the PL spectra of MoS$_2$/TCNQ/WSe$_2$ and MoS$_2$/CoPc/WSe$_2$. The two spectra appear nearly identical, displaying emission peaks at ~666 and ~746 nm corresponding to MoS$_2$ and WSe$_2$, respectively. These spectra are also very similar with those from other heterostructures in Figures 2 and 3. The intralayer excitons may recombine at the valleys of each TMDC layer, where the band energies of one layer are not significantly hybridized with the other layers. Therefore, we conclude that the intralayer emissions from the constituent layers are not significantly affected by the organic layers, which is consistent with our previous study.[2] In contrast, the interlayer emission is not observed from either sample.

Figures 4c and 4d respectively present the energy band diagrams of the TCNQ- and CoPc-inserted heterostructures, illustrating the quenching mechanisms for interlayer emission. The TCNQ LUMO is located lower than both $MoS_2$ and $WSe_2$ CBM. In the same manner, its HOMO is placed below $MoS_2$ and $WSe_2$ VBM, thus forming a discontinuously staggered gap at the interfaces. Due to the alignment, the photoexcited electrons from $MoS_2$ and $WSe_2$ may be trapped in the TCNQ layer. The trapped charges in TCNQ may not be further transferred to other TMDC layers, because there is no energy state in $MoS_2$ and $WSe_2$ near the TCNQ LUMO. As a result, the interlayer exciton formation is suppressed. The CoPc-embedded heterostructure has an opposite band alignment, but displays the same PL characteristics. The LUMO and HOMO levels of CoPc are higher than CBM and VBM of both $MoS_2$ and $WSe_2$, respectively. The alignment now favors trapping of holes and prohibits additional hole transfer from CoPc to $MoS_2$ or $WSe_2$. Therefore, the interlayer excitons may not be formed in $MoS_2$/CoPc/$WSe_2$, thus no related emission is observed. The PL spectra of TCNQ- and CoPc-functionalized individual TMDC flakes in Figure S4 support the trapping behavior of the charges. The TMDC emission intensities are drastically reduced compared to those of the control samples without organic layers. This quenching results from the photoinduced charge transfer at the TMDC/organic interface (see SI for details).

Finally, it is worth noting that the embedded organic layers may not support the radiative recombination of interlayer excitons at the organic/TMDC interfaces. We did not observe any interlayer emission from the hybrid interfaces, for example, $MoS_2$/B3PyPB and $WSe_2$/EY. Instead, only a strong quenching was monitored as shown in Figure S4. This strongly suggests that the Coulomb interaction between the charges separately residing in TMDC and organic layers may not be as strong as that in TMDC/TMDC interfaces. Furthermore, the interlayer emission requires both energy and momentum conservation. The lack of radiative recombination leads us to conclude that the organic layers may not compensate the momentum mismatch with TMDC layers due to the relatively weak coupling between organic and TMDC layers compared to the strong interaction at the $MoS_2$/$WSe_2$ interface.[2]

In closing, we have studied the effects of dielectric properties, separation gap, and band alignments on $X_I$ by inserting various organic layers in $MoS_2$/$WSe_2$ heterostructures. The organic layers decrease the dielectric constant and expands the separation distance between the TMDCs, together reducing the $X_I$ binding energy. This ultimately leads to a strong blueshift of the interlayer emission. Depending on the types of the energy level alignment, interlayer emission can be further blueshifted by the dipolar interaction, or completely quenched due to the trapping of charges. This approach may be further developed to provide additional controllability on interlayer excitons. For example, the inserted layer may be modified to have periodic patterns by using lithographic techniques.[44-45] Such platforms may allow for a study of the localized Coulombic interaction in van der Waals heterostructures[46] and the physics of exciton-polariton condensates[14]. Environmentally sensitive molecules could also be used to tailor interlayer exciton properties by external stimuli, including electro-,[47] chemo-,[48] and photo-chromic[49] molecules. Furthermore, the organic-embedded hybrid heterostructures may be used as a testbed to investigate the interlayer exciton dynamics. The study of organic-TMDC hybrid materials has been limited to characterize the PL lifetime of the intralayer excitons,[50] while the $X_I$ lifetime is not well understood. The $X_I$ lifetime measurement will contribute to our understanding of light-matter interactions. These efforts could lead to a new class of $X_I$-based devices with externally tunable characteristics.

**Materials and Methods**

**Sample fabrication.** Monolayer $MoS_2$ flakes were mechanically exfoliated from bulk crystals (SPI Supplies) and deposited on a p-doped $Si/SiO_2$ substrate. The substrate was ultrasonicated in various solvents, including acetone, methanol, and deionized water (DI) for 30 min each. Then, it was dried by blowing air and placed on a hot plate at 110 °C for 2 min.[22] To fabricate the organic layers on top of the $MoS_2$ flakes, the samples were immersed in the solutions containing the dye molecules at room temperature for 8 h: B3PyPB (1 mg/ml in chloroform), EY (1 mg/ml in ethanol), TCNQ (0.6 mg/ml in chloroform), and CoPc (0.4 mg/ml in dimethylformamide). These molecules are purchased from Sigma-Aldrich in a powder form. The organic-layer/$MoS_2$ samples were subsequently washed with each solvent of the molecules and DI to remove the loosely bound particles. For example, B3PyPB deposited sample was rinsed with the excess amount of chloroform and then cleaned with DI water.

Top $WSe_2$ layer was stacked on the prepared organic/$MoS_2$ sample by a polymer-stamping method.[2] $WSe_2$ flakes were first mechanically exfoliated on a dimethylpolysiloxane (PDMS, purchased from Krayden) stamp. The stamp was fabricated by mixing the base and curing agent of PDMS with 12:1 ratio and cured at 65 °C for 12 h. The stamp with a monolayer of $WSe_2$ was placed on a glass substrate which is connected to a micro-positioner (Signatone). Then, the $WSe_2$ flake was positioned above the $MoS_2$/organic layer and brought into contact. The elastomer maintained the position for 30 min at 50 °C and lifted slowly by manipulating the positioner. Finally, the samples were annealed under argon gas environment at 280 °C for 2 h for a better contact at the interface of the heterolayers.

**Sample characterization.** The optical and Raman measurements were performed using a Renishaw confocal microscope under ambient conditions. A 633-nm HeNe laser at ~0.1 mW was used to irradiate samples through a 100× objective lens. The optical signals of the samples were transferred by the same objective and collected by a charge-coupled device (CCD) camera. The layer numbers of TMDC flakes were first inspected by optical microscopy, and confirmed by the height measurement AFM. A Bruker Dimension Icon AFM was used with a SCANASIST-AIR probe under ambient conditions. To avoid any unwanted effects of the moisture, we baked the samples in the same manner described above before the measurement. The pixel window of $256 \times 256$ was maintained, and the actual size of the window was adjusted depending on the sizes of samples. The raster-scanning was conducted sequentially after each step of stacking to monitor the thickness of each layer in the heterostructure.

**Acknowledgements.**
This work was funded by the U.S. National Science Foundation.

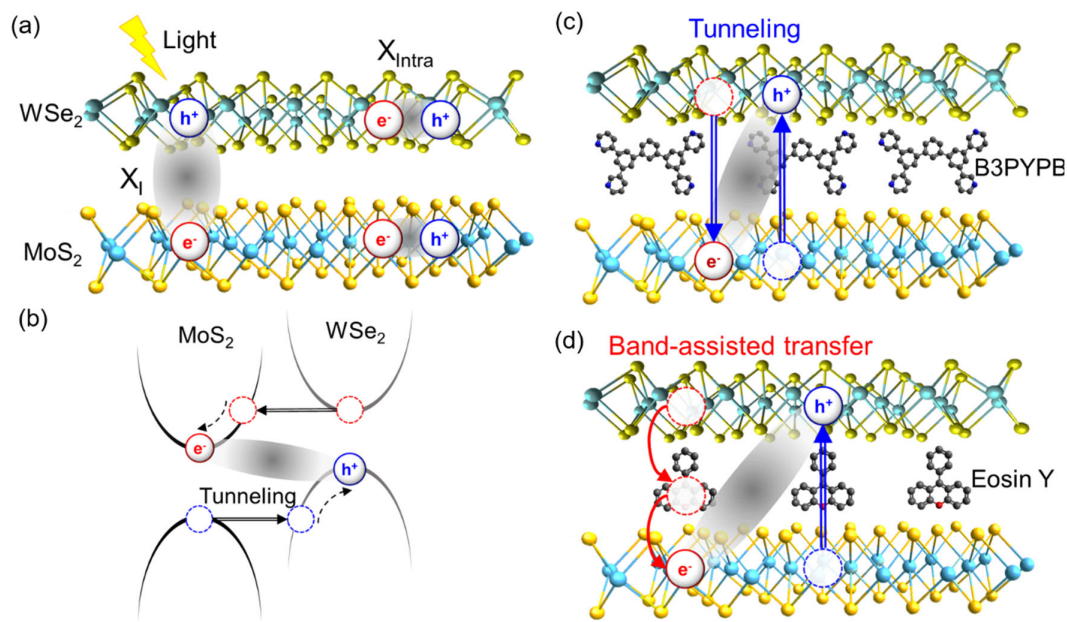

**Figure 1.** (a) Schematic of a MoS$_2$/WSe$_2$ heterostructure. Laser irradiation creates X$_I$ as well as excitons in individual MoS$_2$ and WSe$_2$ layers. The Coulomb interaction between the spatially separate electron-hole pair forms a strong bound state as the interlayer exciton. (b) Mechanism for interlayer emission. The photoexcited electron in WSe$_2$ and hole in MoS$_2$ transfer to the opposite layers via tunneling facilitated by the staggered band alignment. Schematics of the organic layer embedded hybrid heterostructures: (c) MoS$_2$/B3PyPB/WSe$_2$ and (d) MoS$_2$/EY/WSe$_2$. The photoinduced charges in the B3PyPB-embedded heterostructure tunnel through the large-energy B3PyPB layer, while the electron transfer via EY is energetically favorable in MoS$_2$/EY/WSe$_2$.

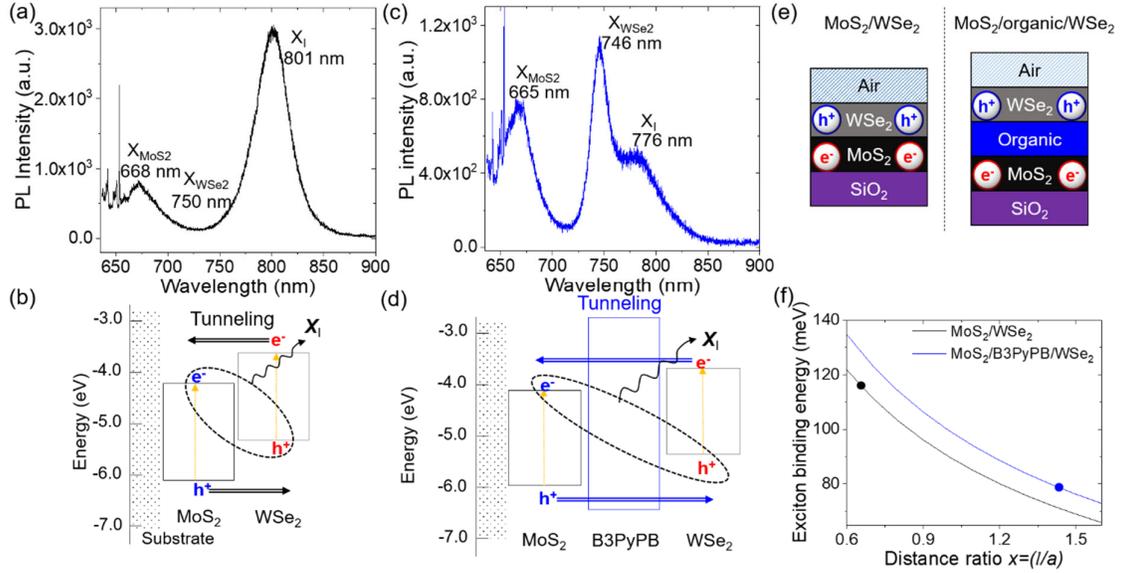

**Figure 2.** (a) PL spectrum of MoS$_2$/WSe$_2$ with a HeNe laser excitation at 633 nm. The deconvolution suggests interlayer emission at ~801 nm as well as PL peaks at ~668 nm (MoS$_2$) and ~750 nm (WSe$_2$). (b) Energy diagram of MoS$_2$/WSe$_2$ and X$_I$ formation in via tunneling of photoinduced charges. (c) PL spectrum of MoS$_2$/B3PyPB/WSe$_2$ showing the interlayer emission with much reduced intensity and a strong blueshift. (d) Energy diagram and schematic of B3PyPB-embedded MoS$_2$/WSe$_2$. The organic layer increases the tunneling length and weakens the dielectric screening, resulting in the 25-nm PL shift. (e) Dielectric model of the control MoS$_2$/WSe$_2$ (left) and hybrid structure (right). The control sample is modelled with four dielectric layers, while the hybrid structure consists of five dielectric layers, including the organic layer. (f) Theoretical X$_I$ binding energies of MoS$_2$/B3PyPB/WSe$_2$ (blue line) and MoS$_2$/WSe$_2$ (black line) as a function of distance ratio $x$. The estimated binding energy difference between the samples (filled black and blue circles, ~50 meV) corresponds to the interlayer emission shift observed with the hybrid heterostructure.

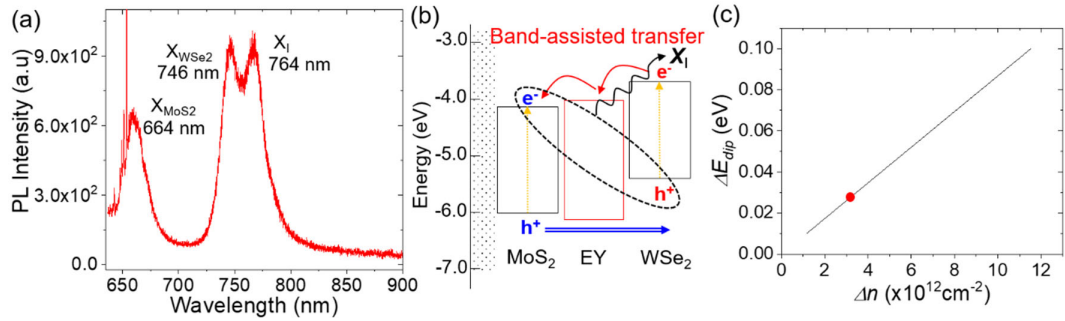

**Figure 3.** (a) PL spectrum and (b) energy band diagram of MoS$_2$/EY/WSe$_2$ heterostructure. Compared to the B3PyPB-inserted sample, the interlayer emission is stronger and blueshifted while intralayer PL signatures are nearly identical. The EY layer facilitates photoinduced electron transfer from WSe$_2$ to MoS$_2$ by providing an energy state between their CBM. The energetically favorable band-assisted transfer will thus lead to greater amounts of electrons and holes (*i.e.*, out-of-plane dipoles) than the B3PyPB-embedded heterostructure. (c) A plot of dipolar interaction energy versus the dipole density. The interaction energy linearly increases with the dipole density.

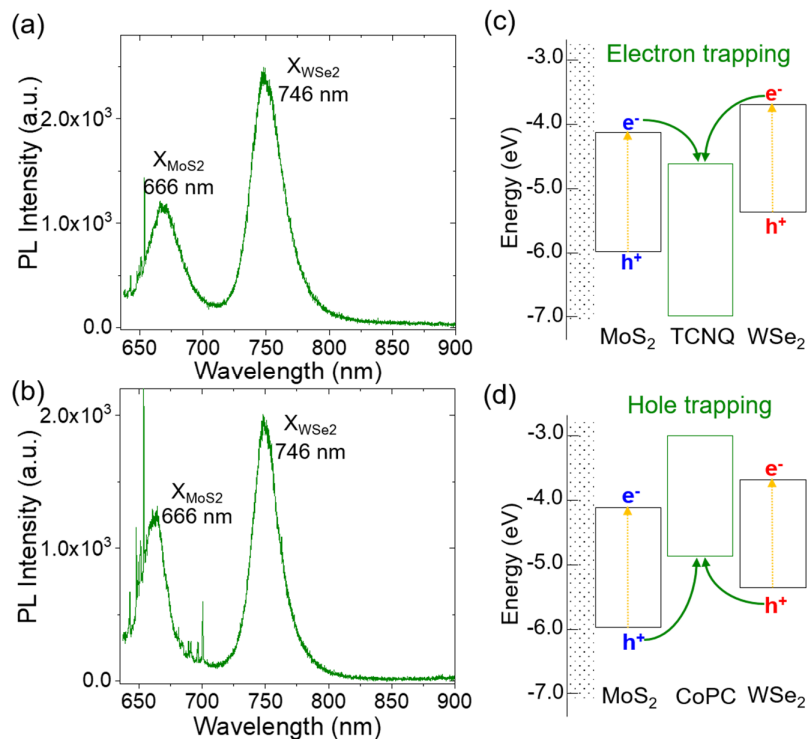

**Figure 4.** PL spectra of hybrid heterostructures including a layer of (a) TCNQ and (b) CoPc between TMDCs. The intralayer PL spectra are nearly identical, whereas the interlayer emission is not observed from both hybrid heterostructures. Energy band diagrams of (c) $MoS_2$/TCNQ/$WSe_2$, favoring electron trapping from both TMDCs, and (d) $MoS_2$/CoPc/$WSe_2$, supporting the hole trapping. Both trapping processes induce the complete quenching of $X_I$ emission.

Supporting Information

# Understanding the Effects of Dielectric Property, Separation Distance, and Band Alignment on Interlayer Excitons in 2D Hybrid MoS$_2$/WSe$_2$ Heterostructures

Jaehoon Ji and Jong Hyun Choi*
School of Mechanical Engineering, Purdue University
West Lafayette, Indiana 47907, United States
*Corresponding author: jchoi@purdue.edu

**Content**
1. AFM Images and Height Profiles of MoS$_2$/organic/WSe$_2$ Hybrid Heterostructures
2. Deconvoluted PL Spectra of MoS$_2$/organic/WSe$_2$ Hybrid Heterostructures
3. AFM and PL Measurements of TMDC/organic Samples
4. Solution for 2D Hydrogen-like Schrödinger Problem
5. Dielectric Model
6. References

## 1. AFM Images and Height Profiles of MoS$_2$/organic/WSe$_2$ Hybrid Heterostructures

To characterize each constituent layer in 2D van der Waals structures, direct imaging techniques such as AFM imaging and transmission electron microscopy (TEM) have been adopted in numerous studies.[1-6] Among the characterization methods, we used AFM imaging to show the layer number of TDMC flakes and the presence of the organic layers in between the TMDC layers. AFM measurement does not damage the crystal structure of the 2D van der Waals heterostructures[7] and is ideally suited to monitor the hybrid samples. Figure S1 shows the raster-scanned AFM images and the surface profiles of MoS$_2$/organic/WSe$_2$ samples and the control MoS$_2$/WSe$_2$. All MoS$_2$/organic/WSe$_2$ hybrid structures have a uniform thickness around ~2.9 nm, which is consistently higher than that of MoS$_2$/WSe$_2$ by ~0.9 nm. This indicates that the organic molecules assemble as a uniform 2D layer between TMDC monolayers.

TEM cross-sectional imaging of the samples may be another option for the characterization. However, one of the inevitable steps in TEM imaging is the slicing of the samples by ion-beam milling or etching the unwanted area.[1-2] Under the harsh preparation conditions, the hybrid samples may not preserve enough size to distinguish each constituent layer. Moreover, the organic layers in our samples will be vulnerable under the harsh conditions and could be damaged during the preparation.[8] Therefore, we used AFM and PL measurements (instead of TEM) to confirm the presence of the organic layers. Specifically, we measured additional AFM images of individual flakes with and without the organic layers, and their corresponding PL spectra (see Section 3 below).

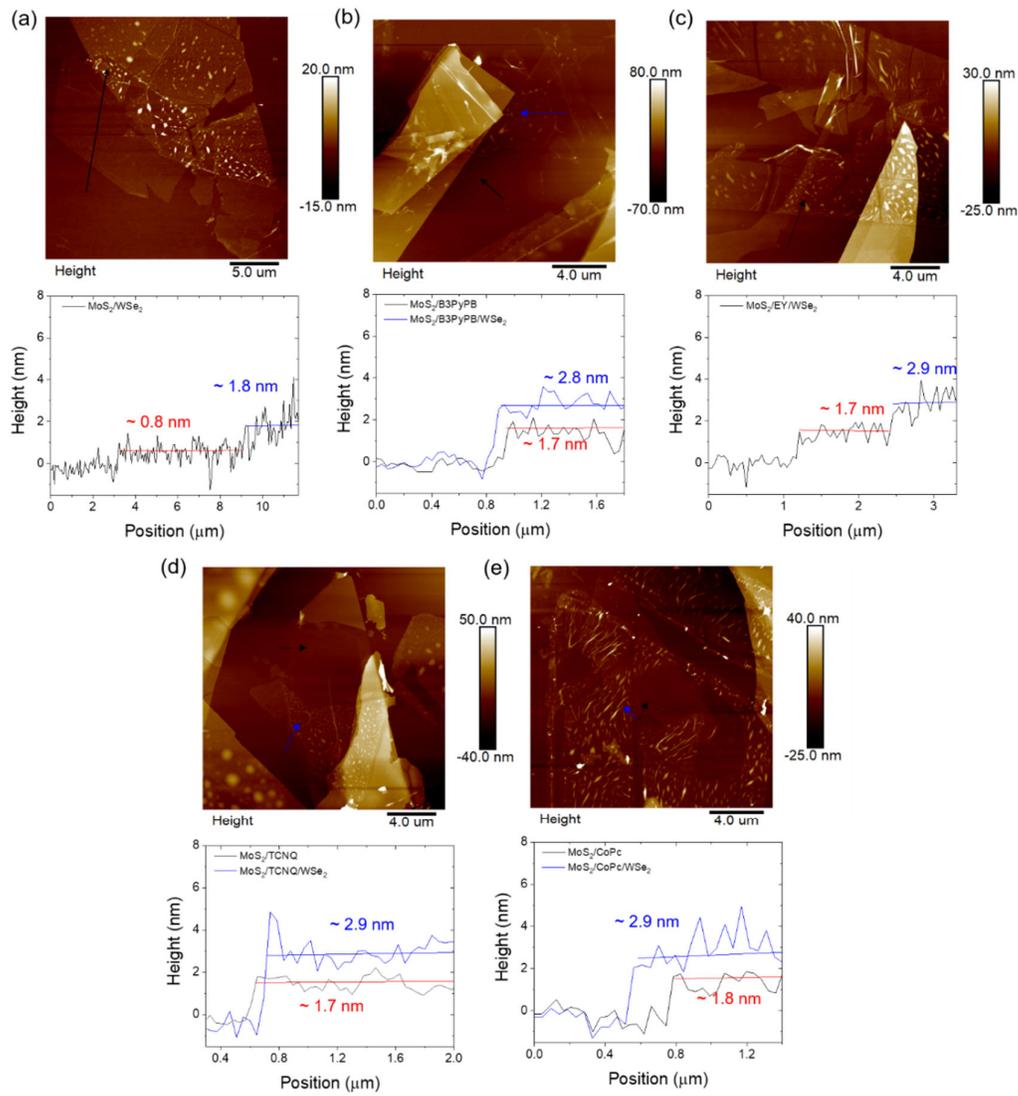

**Figure S1.** Raster-scanned AFM images and height profiles of (a) $MoS_2/WSe_2$ (control), (b) B3PyPB-, (c) EY-, (d) TCNQ-, and (e) CoPc-embedded hybrid-structures, where the organic layers are sandwiched by the TMDC flakes. The height profiles are measured along the arrow lines in the height map.

## 2. Deconvoluted PL Spectra of MoS$_2$/organic/WSe$_2$ Hybrid Heterostructures

The steady-state PL spectra of the samples were measured under the light irradiation with a 633 nm HeNe laser. The temperature-dependent PL spectroscopy may provide insights into the characteristics of interlayer excitons from hybrid heterostructures. However, we did not perform temperature-dependent measurement in this work due to the stability of the organic layers. For example, TCNQ molecules show a phase transition at temperature near 200 K.[9] The phase transition could deteriorate the quality of the organic layer may deteriorate and induce unwanted effects on the excitons.

To eliminate any temperature-dependent uncertainties, the PL spectra of all the samples were measured at room temperature. Figure S2 shows that intralayer PL emission peaks of MoS$_2$ and WSe$_2$ are approximately 666 nm and 746 nm, respectively, for all hybrid heterostructures. The interlayer emission, on the other hand, differs for various organic layers. The control MoS$_2$/WSe$_2$ sample exhibits the interlayer emission at ~801 nm; MoS$_2$/B3PyPB/WSe$_2$ has $X_I$ peak at ~776 nm; the interlayer emission from MoS$_2$/EY/WSe$_2$ is at ~764 nm. No interlayer emission is observed from MoS$_2$/TCNQ/WSe$_2$ and MoS$_2$/CoPc/WSe$_2$. The detailed mechanisms for the observed phenomena are discussed in the main text.

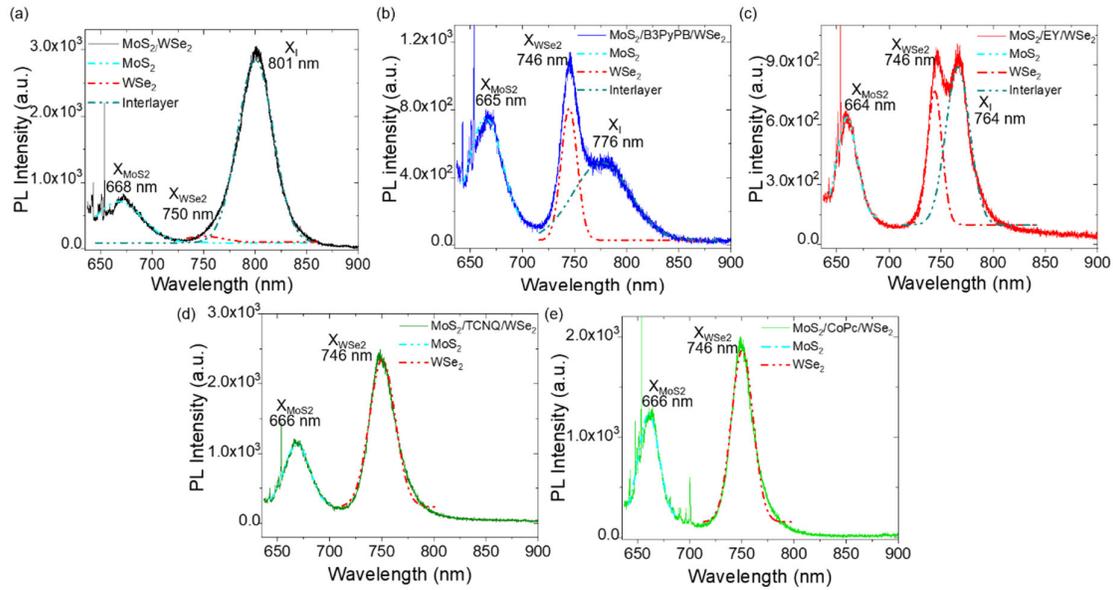

**Figure S2.** PL spectra of (a) MoS$_2$/WSe$_2$ (control), (b) B3PyPB-, (c) EY-, (d) TCNQ-, and (e) CoPc-embedded heterostructures. Three Gaussian profiles are used to deconvolute each PL spectrum: MoS$_2$ (cyan) and WSe$_2$ (red), and interlayer emission (dark cyan). The wavelengths of the deconvoluted emission peaks are noted in each figure.

## 3. AFM and PL Measurements of TMDC/organic Samples

We verified the presence of organic layers on individual TMDC monolayers. The samples were prepared by mechanically exfoliating monolayer TMDC flakes on Si substrates and immersing them in the solutions containing organic molecules (see Materials and Methods for sample preparation details). We measured the thicknesses of $MoS_2$/B3PyPB and $MoS_2$/CoPc samples with AFM as shown in Figure S3. The AFM imaging of other TMDC/organic samples ($WSe_2$/EY and $WSe_2$/TCNQ) was previously reported elsewhere.[5-6] The hybrid structures exhibit a uniform height (~1.7 nm) which is higher than that of the monolayer TMDCs by ~0.9 nm. Note that this height difference is identical with the thickness of the organic layer sandwiched by $MoS_2$ and $WSe_2$ in Figure S1.

We also monitor the PL spectra of the individual TMDC monolayers with and without organic layers to investigate the effects of organic layers on individual flakes. Figure S4 shows that the PL intensities of all the TMDC flakes are drastically quenched by the organic layers. We attribute this to the interplay between (i) dark-state doping and (ii) photo-induced charge transfer. (i) As the work function of TMDCs and the reduction potential of organic layers become aligned at the interface, the charge transfer may occur at the interface regardless of the illumination. For example, the reduction potential of B3PyPB (~-2.8 eV) is higher than the work function of $MoS_2$ (~-4.6 eV).[10] To balance the difference, B3PyPB may donate electrons to $MoS_2$ (*i.e.*, n-doping), resulting in the reduction of $MoS_2$ PL (Figure S4a). (ii) Photo-induced charge transfer may also take place at the TMDC/organic interface. For example, $WSe_2$ CBM is higher than EY LUMO as illustrated in Figure 3b. As a result, photo-induced electrons will migrate from $WSe_2$ CBM to EY LUMO, thereby quenching the PL of $WSe_2$ flakes (Figure S4b). Similarly, $WSe_2$ VBM is located lower than CoPc HOMO (Figure 4d). Due to the built-in potential at the $WSe_2$/CoPc interface, the holes in the $WSe_2$ VBM will transition to CoPc MOHO, resulting in a quenching of $WSe_2$ PL (Figure S4d). More detailed charge transfer mechanisms were discussed in our previous publications.[5-6] The AFM images and PL spectra together confirm that uniform 2D organic layers are formed on TMDC flakes, which drastically modulate the PL emission properties of TMDCs.

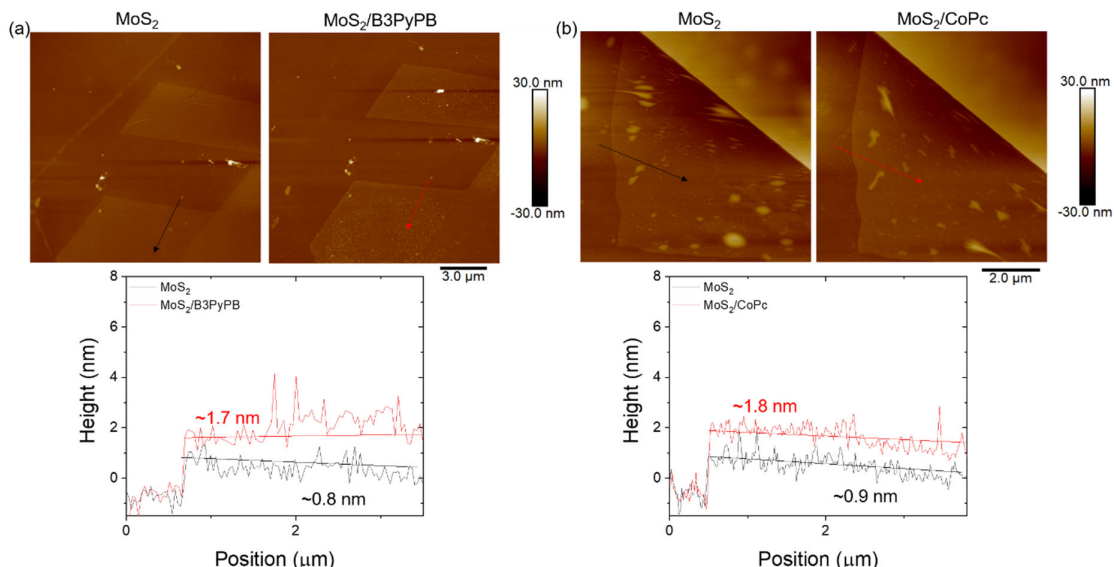

**Figure S3** Raster-scanned AFM images and height profiles of (a) $MoS_2$/B3PyPB and (b) $MoS_2$/CoPc. The height of individual TMDC monolayers is about 0.8 nm. Compared with the pristine TMDC monolayers, the organic-layer-deposited TMDC flakes have consistently

greater heights by ~0.9 nm. This height increase is consistent with the height difference between $MoS_2$/$WSe_2$ and $MoS_2$/organic/$WSe_2$ (see section 1). The AFM measurements of EY- and TCNQ-deposited TMDC samples were investigated in our previous work.[6]

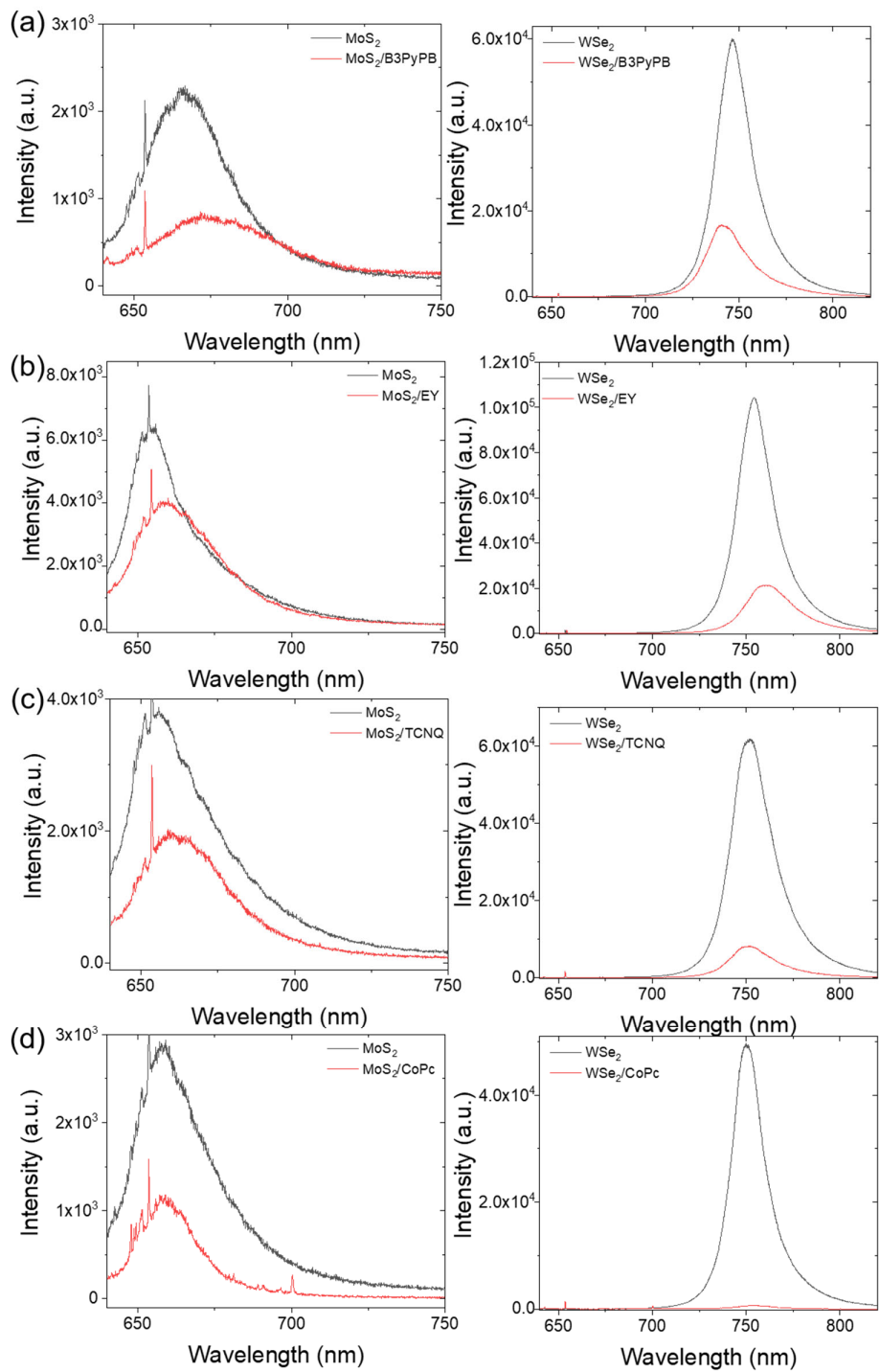

**Figure S4.** PL spectra of $MoS_2$ (left) and $WSe_2$ (right) samples functionalized with (a) B3PyPB, (b) EY, (c) TCNQ, and (d) CoPc layers. All the organic-functionalized TMDC flakes emit significantly weaker PL than the control samples without the organic layers, but the extent of the modulation depends on the combinations of TMDCs and organic layers. The PL quenching may be attributed to the dark-state doping and the photo-induced charge transfer at the TMDC/organic interface. More detailed discussion on the effects of the organic-functionalization is presented in our previous reports.[5-6]

## 4. Solution of 2D Hydrogen-like Schrödinger Problem

An exciton is a quasi-particle formed as a bound state of photoexcited electron and hole, similar to the hydrogen atom with the electrostatic interaction between the constituent charges.[11] From the Wannier equation[12], the energy of the particle can be expressed as a sum of the kinetic energy from the relative motion of the charges and the electrostatic interaction within the particle. The eigenvalue of the energy (i.e., exciton binding energy or $E_b$) can be obtained by solving a Schrödinger equation of the hydrogen-like particle[13-14]:

$$-\frac{\hbar}{2m_{ex}}\frac{1}{\rho}\frac{d}{d\rho}\left[\rho\frac{d\varphi(\rho,l)}{d\rho}\right] - \frac{e^2}{\varepsilon_{ex}\sqrt{\rho^2+l^2}}\varphi(\rho,l) = E_b\varphi(\rho,l) \quad . \quad (1)$$

where $\hbar$ is the reduced Planck constant ($\hbar = h/2\pi$), $e$ is the electrical charge, $\rho$ is the relative coordinate, $m_{ex}$ denotes the effective mass of the exciton, and $l$ is the separation distance between photoexcited electron and hole residing in $MoS_2$ and $WSe_2$, respectively. $\varepsilon_{ex}$ is the dielectric constant of the exciton.[15-17] $\varphi$ and $E_b$ are the eigenfunction and eigenvalue, respectively. To simplify the equation, we define the radius of the interlayer exciton as $a = \varepsilon_{ex}\hbar/m_{ex}e^2$ and $E_{b0}$ as the binding energy of $X_I$ without the separation given by $l$: $E_{b0} = m_{ex}e^4/2\varepsilon_{ex}\hbar^2$.[18] The charge separation may be determined by considering the thicknesses of TMDC layers and the intermediate layer.[14] The hybrid structure has the intermediate organic layer with a thickness of ~0.9 nm and the separation distance of $l = $ ~1.7 nm. The $l$ of the control sample without the intermediate layer is set to be ~0.8 nm.

From the effective dielectric constant of each structure (see Section 4 below), we obtain $E_{b0} = $ ~114 meV for the hybrid structure and $E_{b0} = $ ~103 meV for the control ($MoS_2/WSe_2$), respectively. The $E_{b0}$ value of the control sample is in excellent agreement with previous reports.[19] From the relationship, the equation (1) may be converted into a dimensionless form:

$$-\frac{1}{y}\frac{d}{dy}\left[y\frac{d\psi}{dy}\right] - \frac{2}{\sqrt{y^2+x^2}}\psi = w(x)\psi \quad (2)$$

where $y = \rho/a$, $x = l/a$, $w(x) = E_b/E_{b0}$, and $\psi = \varphi/a$. The solution of the differential equation can be reduced to the function derived from the curve-fitting method reported by Leavitt et al.[13]:

$$w(x) = \frac{4+12.97x+0.718x^2}{1+9.65x+9.24x^2+0.3706x^3} \quad (3)$$

Figure S3 displays the binding energy of the interlayer exciton as a function of the distance ratio $x$. $E_b$ is inversely proportional to $x$. As the ratio increases, the exciton binding energy monotonically drops. It implies that with a larger separation gap, the overlap between spatially separate electron and hole becomes weaker, thereby resulting in a smaller binding energy.

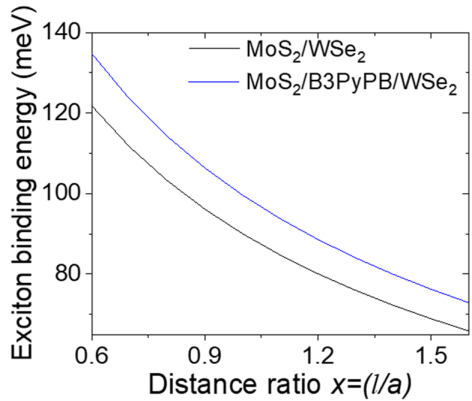

**Figure S5.** Theoretical prediction of the interlayer exciton binding energies in $MoS_2/WSe_2$ (black) and B3PyPB-inserted heterostructure (blue) as a function of distance ratio $x$. The $X_I$ binding energy decreases monotonically with increasing distance ratio.

## 5. Dielectric Model

The Coulomb interaction between the charges in $X_I$ highly depends on the dielectric screening of the quasi-particle. To account for the dielectric properties within the heterostructure, a dielectric model is used with a multi-layer Poisson's equation.[15] As illustrated in Figure 2e, five-layers and four-layers of piecewise dielectrics are modelled to reflect the dielectric properties of the organic-inserted structure and the control sample (without an organic layer), respectively. $\varepsilon_i$ in Figure 2e represents the relative dielectric constant of the $i_{th}$ layer of the heterostructure. The bottom layer $SiO_2$ has a dielectric constant of ~3.9, while the top layer, air, has the constant near 1. Both TMDC layers have similar dielectric properties (~ 9.7 for $MoS_2$ and ~10.6 for $WSe_2$).[14] The intermediate organic layer between TMDCs has a dielectric constant of ~ 3.[15, 20-22] For charges in $z$ and $z'$ planes, the electrostatic potential ($\Phi$) with in-plane coordinate $\rho$ can be expressed as:[14]

$$-V(\rho, z, z') = \frac{1}{4\pi^2}\int \Phi(z, z', k)e^{ik\rho}d^2k, \qquad (4)$$

where the potential $V$ satisfies the Poisson's equation and $k$ is the inverse of the distance of the charges.

$$4\pi\delta(z-z') = [k^2\varepsilon(z,k) - \frac{\partial}{\partial z}\varepsilon(z,k)\frac{\partial}{\partial z}]\Phi(z,z',k). \qquad (5)$$

As the charges of $X_I$ are confined in TMDC layers, the potential ($\Phi_{TMDC}$) from $X_{IS}$ has a relationship:

$$\Phi_{TMDC}(z,z',k) = \varepsilon_{ex}\Phi(z,z',k). \qquad (6)$$

From the five-layer model reported by Chen et al.[23-24], the effective dielectric constant ($\varepsilon_{ex}$) of the hybrid structure is estimated to be ~5.4. The control TMDC heterostructure has $\varepsilon_{ex}$ = ~5.7 which is obtained by using a four-layer model presented by Kamban et al.[14]